\documentclass[conference,10pt]{IEEEtran}
\IEEEoverridecommandlockouts

\usepackage{cite}
\usepackage{graphicx}
\usepackage{amsmath}
\usepackage{array}
\usepackage{mdwmath}
\usepackage{amssymb}
\usepackage{mdwtab}
\usepackage{stfloats}
\usepackage[tight,footnotesize]{subfigure}
\usepackage{amsmath,amsthm}
\usepackage{threeparttable}
\usepackage{color}
\usepackage{url}
\usepackage{algpseudocode}
\usepackage{algorithm}
\usepackage{multicol}
\usepackage{amssymb}

\algrenewcommand{\algorithmicrequire}{\textbf{Input:}}
\algrenewcommand{\algorithmicensure}{\textbf{Output:}}

\def\BibTeX{{\rm B\kern-.05em{\sc i\kern-.025em b}\kern-.08em
    T\kern-.1667em\lower.7ex\hbox{E}\kern-.125emX}}

\begin{document}

\title{Performance Analysis of RIS-Aided Space Shift Keying With Channel Estimation Errors}

\author{
	\IEEEauthorblockN{
       Xusheng Zhu, Wen Chen, Qingqing Wu, Liwei Wang
}
	\IEEEauthorblockA{Department of Electronic Engineering, Shanghai Jiao Tong University, Shanghai, China}
	\IEEEauthorblockA{Email: \{xushengzhu, wenchen, qingqingwu, wanglw2000\}@sjtu.edu.cn}
}

\markboth{}
{}
\maketitle

\begin{abstract}
In this paper, we investigate the reconfigurable intelligent surface (RIS) assisted space shift keying (SSK) downlink communication systems under the imperfect channel state information (CSI), where the channel between the base station to RIS follows the Rayleigh fading, while the channel between the RIS to user equipment obeys the Rician fading.
Based on the maximum likelihood detector, the conditional pairwise error probability of the composite channel is derived. Then, the probability density function for a non-central chi-square distribution with one degree of freedom is derived. Based on this, the closed-form analytical expression of the RIS-SSK scheme with imperfect CSI is derived.
To gain more valuable insights, the asymptotic ABEP expression is also given.
Finally, we validate the derived closed-form and asymptotic expressions by Monte Carlo simulations.
\end{abstract}
\begin{IEEEkeywords}
Reconfigurable intelligent surface, space shift keying, imperfect channel state information, average bit error probability.
\end{IEEEkeywords}

\section{Introduction}
Reconfigurable intelligent surface (RIS) has recently attracted considerable concern thanks to their ability to make environments controllable \cite{wu2019towards}.
Particularly, RIS is an electromagnetic metasurface comprising small, low-cost, and almost passive scattering elements that can induce a predetermined phase shift in the incident wave \cite{Taosum2021}. Consequently, RIS can efficiently modify the scattering, reflection, and refraction of the environment cost-effectively, thereby improving the efficiency of wireless networks \cite{saad2020vis}.
To clarify the impact of multiple RIS on system performance, \cite{do2021multi} investigated the statistical characteristics and modeling of distributed multiple RIS-assisted wireless systems.
In addition, \cite{tang2021wireless} conducted measurements of the path loss of RIS-assisted wireless communication in a microwave radio chamber taking into account different scenarios.

Spatial modulation (SM), or more generally index modulation (IM), has recently gained significant research attention due to its efficient energy utilization \cite{zhu2021performance}.
In particular, IM offers a solution to this issue by utilizing an index of available resources, such as transmit or receive antennas and frequency domain subcarrier, to convey a portion of the information \cite{li2023index}.
This allows only a fraction of the energy-consuming resources to be activated at any given time, making IM a highly energy-efficient option. For this reason, IM is viewed as a promising technology for 6G systems \cite{li2023index}.
With the aim of focusing more on spatial domain information, \cite{jegan2009space} studied the space shift keying (SSK) scheme by neglecting the symbol domain information of the SM.
Due to the large path loss in the millimeter wave (mmWave) band, it is difficult to guarantee the reliability of the received data by utilizing SM techniques for signal transmission at each time slot. In this regard, \cite{zhu2023qua} proposed a new quadrature spatial scattering modulation scheme that exploits the hybrid beamforming instead of a single antenna in the SM.

In light of the advantages possessed by RIS and SSK, the RIS-assisted SSK scheme has attracted extensive research interest from the academic community
\cite{can2020re,canbilen2022on,li2021space}.
Specifically, the RIS-aided SSK scheme was presented in \cite{can2020re}, where the antenna is switched and selected at the transmitter side and the RIS is viewed as a passive relay.
In \cite{li2021space}, the RIS incorporates Alamouti space-time block coding, allowing the RIS to send its Alamouti-encoded data and reflect the incoming SSK signals toward the target.
Moreover, \cite{canbilen2022on} takes the case of a hardware-impaired transceiver into account and analyzes its impact on the average bit error probability (ABEP) of the RIS-SSK scheme.
With the aim of studying RIS for mmWave information transmission, \cite{zhu2021ris} proposed a RIS-assisted spatial scattering modulation scheme and provides a theoretical analysis with respect to ABEP.

All the above-mentioned RIS-aided literature assumes that the channel state information (CSI) is completely well-known at the transceiver. Nevertheless, in reality, the estimated CSI is imperfect on account of estimation errors and limited radio resources of the RIS.
Although work on RIS-assisted communication systems under imperfect CSI is common, there has been no work on RIS-assisted IM schemes under imperfect CSI in the open literature.
Against this background, we intend to elucidate this timely and interesting topic.
To the best of our knowledge, there is no analytical approach that has been adopted to investigate imperfect CSI for RIS-assisted SSK system error performance.
For clarity, the contribution of this paper are summarized as follows:
1) In this paper, we study RIS-assisted SSK systems, where the base station to RIS (BS-RIS) channel obeys the Rayleigh fading, while the RIS to user equipment (RIS-UE) channel obeys the Rician fading. We consider that perfect CSI estimation can be obtained in the BS-RIS channel, while perfect CSI cannot be estimated in the RIS-UE channel.
2) The maximum likelihood (ML) detection algorithm is used to adjudicate the RIS-SSK scheme and derive the conditional pairwise error probability (CPEP) expressions. In addition, we provide the complexity analysis. By utilizing the central limit theorem (CLT), we derive the expectation and variance of the composite channel. After that, we derive the probability density function (PDF) from the BS to UE.
3) Based on the derived CPEP and PDF, we derive the closed-form solution for the unconditional pairwise error probability (UPEP) considering the imperfect CSI case. Further, asymptotic UPEP and ABEP expressions are both derived. We have exhaustively verified the ABEP expressions via the Monte Carlo simulations.

\begin{table}[t]
\small
\begin{tabular}{ll}
\hline Notations & Definitions \\
\hline$N_{\mathrm{t}}$ & Number of transmit antenna \\
$N_{\mathrm{r}}$ & Number of receive antenna \\
$n_{\mathrm{t}}$ & Transmit antenna index \\
$P_s$ & The average transmit power \\
$\mathbb{C}^{m\times n}$ & The space of $m\times n$ matrics\\
$|\cdot|$ & The absolute value operation \\
$\lambda$ & Wavelength \\
${\rm diag}(\cdot)$ & The diagonal matrix operation \\
$\mathbb{C}^{n\times m}$ & The space of $n\times m$ complex-valued matrices \\
$(\cdot)^T$ & The transpose operator \\
$I_0(\cdot)$ & The first kind of zero-order modified Bessel function\\
$I_1(\cdot)$ & The first kind of first-order modified Bessel function\\
$\mathcal{N}(\cdot,\cdot)$ & The real Gaussian distribution\\
$\mathcal{CN}(\cdot,\cdot)$ & The circularly symmetric complex \\& Gaussian distribution \\
$\Re\{\cdot\}$ &  The real part of a complex variable \\
$\Pr(\cdot)$ & The probability of an event \\
$P_b$ & CPEP \\
$\bar P_b$ & UPEP\\
$\sim$ & ``Distributed as"\\
$\mathrm{E}(\cdot)$ & The expectation operation \\
$Var(\cdot)$ & The variance operator\\
$Q(\cdot)$ & The Q-function  \\
\hline
\end{tabular}
\end{table}
\section{System Model}
In this section, we study the RIS-SSK system model under imperfect CSI, where optimal reflection phase of RIS is considered.
It is assumed that the SSK technique is used by mapping the input bits to the index of a specific transmit antenna, which is activated to allow the transmit signal to reach the UE through the RIS.
The fading channel between the $n_t$-th transmit antenna
and the $l$-th reflective element of the RIS is represented by $g_{l,{n_t}}=\alpha_{l,n_t}e^{- j\theta_{l,n_t}}, n_t\in {1,\cdots, N_t}$.
Meanwhile, the channel between the $l$-th RIS reflecting shift and the receive antenna is denoted by $h_l=\beta_le^{-j\psi_l}$, $l = 1,\cdots,L$.
It is worth noting that the RIS controller can adjust the reflected phase shift to maximize the SNR by adjusting the reflecting phase shift based on the acquired CSI.
In particular, it is assumed that the direct link between BS and UE is not reachable due to undesirable channel conditions and that communication occurs only through the RIS.

Let us consider the communication system shown in Fig. \ref{sysim} that utilizes a RIS to assist the communication between BS to the UE, where the BS consists of $N_{\rm t}$ transmit antennas, while the UE is equipped with a single antenna \cite{can2020re}.
Besides, the  RIS is comprised of a dimensional uniform linear array (ULA) with $L$ reflective elements.
Due to the existence of blockage, RIS is deployed to assist the communication between {BS} and {UE}, where the RIS is fixed to the exterior wall of the building, thus enabling accurate estimation of the indirect channel by calculating the slowly changing arrival and departure angles \cite{zhou2020robust}.
In contrast, reflective channels are more challenging to acquire as the location of the user and environmental factors change.
Considering this, we suppose that the BS-RIS link is perfect, while the RIS-UE link is imperfect owing to channel estimation errors \cite{yang2022per}.
\subsection{Channel Model}
\begin{figure}[t]
  \centering
  \includegraphics[width=7cm]{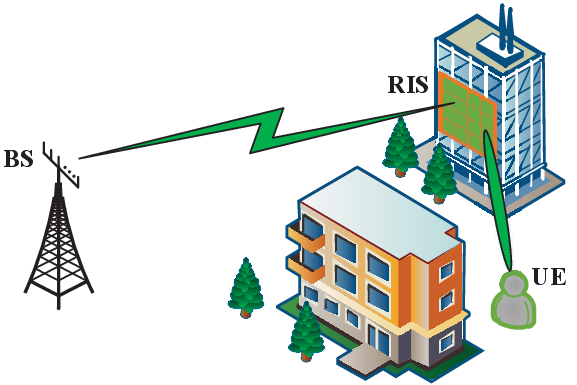}\\
  \caption{\small{System model of the RIS-SSK scheme.}}\label{systemmod}
  \label{sysim}
\end{figure}

In Fig. \ref{sysim}, the RIS-UE channel $\mathbf{h}\in \mathbb{C}^{1\times L}$ is modeled as Rician fading channel, which can be characterized as
\begin{equation}
\mathbf{h} = \zeta \mathbf{\hat h} +\sqrt{1-\zeta^2}\Delta\mathbf{h},
\end{equation}
where $\zeta$ is the correlation coefficient between $\mathbf{h}$ and $\mathbf{\hat h}$, where $\mathbf{\hat h}$ is the information obtained by the channel estimation technique, and $\mathbf{h}$ stands for the practical channel obtained at the UE side. The corresponding estimation error is denoted by $\Delta\mathbf{h}$.
In particular, $\mathbf{\hat h}$ and $\Delta\mathbf{h}$ are mutually uncorrelated.
On the other hand, the BS-RIS channel $\mathbf{g}_{n_t}$ can be modeled as Rayleigh fading channel with non-line-of-sight (NLoS) components.
For the RIS, we set the amplitude of each reflection element of the RIS to one \cite{xx2007tab}.
Based on this, the reflection matrix of the RIS is modeled as
\begin{equation}\label{phaseshif}
\boldsymbol{\Phi} = {\rm diag}(e^{j\phi_{1,n_t}},\cdots,e^{j\phi_{l,n_t}},\cdots,e^{j\phi_{L,n_t}}),
\end{equation}
where $e^{j\phi_{l,n_t}}$ denotes the phase shift that is related to the RIS controller connected to the $n_t$-th activated transmit antenna and the $l$-th reflecting element.

The reflection estimation channel $\mathbf{h}$ can be formulated as \cite{yang2022per}
\begin{equation}\label{h1}
\mathbf{\hat h}=\sqrt{\frac{\kappa}{\kappa+1}}\mathbf{\hat h}^{LoS}+\sqrt{\frac{1}{\kappa+1}}\mathbf{\hat h}^{NLoS},
\end{equation}
where the deterministic line-of-sight (LoS) path matrix can be modeled as
\begin{equation}
\begin{aligned}
\mathbf{\hat h}^{LoS}(\varphi) =& [1,e^{j\frac{2\pi d}{\lambda}\sin\varphi},\cdots,e^{j\frac{2\pi d}{\lambda}l\sin\varphi},\\
&\cdots,e^{j\frac{2\pi d}{\lambda}(L-1)\sin\varphi}]^T,
\end{aligned}
\end{equation}
where $l$ represents the indices of the RIS element.
Besides, the expectation and variance of the magnitude of the path from the $l$-th reflecting element to the UE can be respectively expressed as \cite{san2007dig}
\begin{subequations}\label{beta}
\begin{align}
&E(\hat \beta_l)=\sqrt{\frac{\pi}{4\kappa+4}}e^{-\frac{\kappa}{2}}\left[(1+\kappa)
I_0\left(\frac{\kappa}{2}\right)+\kappa I_1\left(\frac{\kappa}{2}\right)\right],\\
&Var(\hat \beta_l) = 1-{ E}^2(\hat \beta_{l}),
\end{align}
\end{subequations}
where each component of $\mathbf{\hat h}^{NLoS}$ suffers from $\mathcal{CN}(0,1)$.

On the other hand, the BS-RIS channel can be indicated as
$\mathbf{g}_{n_t}\sim \mathcal{CN}(0,\mathbf{I}_{L\times L})$, that is, the path from the activated antenna to the $l$-th reflecting element of the RIS can be written as
$g_{l,n_t}\sim\mathcal{CN}(0,1)$.
Accordingly, the mean and variance of the magnitude on the $l$-th reflecting element to $n_t$-th transmit antenna can be evaluated as \cite{san2007dig}
\begin{equation}\label{alpha}
\begin{aligned}
&E(\alpha_{n_t,l})={\sqrt{\pi}}/{2}, \ \ \ Var(\alpha_{n_t,l}) = (4-\pi)/{4}.
\end{aligned}
\end{equation}
At the UE side, the received signal can be given as
\begin{equation}\label{y01}
y=\sqrt{P_s}\mathbf{h}\boldsymbol{\Phi}\mathbf{g}_{n_t}x  + n_0,
\end{equation}
where $n_0\sim \mathcal{CN}(0,N_0)$ stands for the additive white Gaussian noise (AWGN).
Note that $x$ denotes the Gaussian data symbol, which is a random variable with zero mean and unit variance satisfying $E(|x|^2)=1$.
In this scheme, we aim to study the RIS-aided SSK technique, so the $x$ term can be neglected.
Consequently, the (\ref{y01}) can be re-expressed as
\begin{equation}\label{y02}
y=\sqrt{P_s}\sum\nolimits_{l=1}^Lh_le^{j\phi_{l,n_t}}g_{l,n_t}  + n_0,
\end{equation}
where $h_l = \zeta {\hat h} +\sqrt{1-\zeta^2}\Delta{h}$.
Further, the (\ref{y02}) can be written as
\begin{equation}\label{y03}
\begin{aligned}
y=&\sqrt{P_s}\zeta\sum_{l=1}^L\hat h_le^{j\phi_{l,n_t}}g_{l,n_t}+ \\& \sqrt{P_s(1-\zeta^2)}\sum_{l=1}^L\Delta{h}e^{j\phi_{l,n_t}}g_{l,n_t} + n_0,
\end{aligned}
\end{equation}
where $\Delta h$ represents the error of channel estimation $\hat h$ of and obeys $\mathcal{CN}(0,\sigma_e^2)$ distribution. Particularly, $\sigma_e^2$ represents the variance of the estimation error, which depends on the estimation strategy and the number of pilot symbols employed \footnote{It is worth noting that $\sigma_e^2$ denotes the several factor on the CSI due to limited feedback and channel estimation.
Even in the high SNR region, the channel obtained at the UE is still inaccurate.}.
By adopting orthogonal pilot channel estimation sequences, the estimation error decreases linearly with the increase in the number of pilots.
According to \cite{basar2012per}, the correlation coefficient can be set as $\zeta = 1/\sqrt{1+\sigma_e^2}$.
It is worth mentioning that when $\sigma_e^2=0$, $\zeta = 1$ can be obtained, which indicates perfect channel estimation.
The RIS can adjust the phase shift to make $\phi_{l,n_t}=\theta_{l,n_t}+\psi_l$, thus maximizing the energy of the desired signal of the UE.
\subsection{Detector and Complexity}
\subsubsection{Detector}
In this manner, the received signal can be demodulated by the ML detector, which can be given by
\begin{equation}\label{intdec}
[{\hat n_t}]=\arg\min\limits_{n_t\in\{1,\cdots,N_t\}}\left|y-\sqrt{P_s}\zeta\sum\nolimits_{l=1}^L\alpha_{l,{n_t}}\hat\beta_l\right|^2.
\end{equation}
\subsubsection{Complexity Analysis}
Note that every complex multiplication requires 4 real multiplications and 2 real additions. Computing the square of the absolute value of a complex number requires 2 real multiplications and 1 real addition. In (\ref{intdec}), computing $\sum_{l=1}^{L}\alpha_{l,n_t}\beta_l$ requires $L$ real multiplications and $(L-1)$ real additions.
Computing $\sqrt{P_s}$ and $\zeta$ requires 2 real multiplications.
Subtracting $\sum_{l=1}^{L}\alpha_{l,n_t}\beta_l$ from $y$ requires 1 real addition. At this point, with $L+2$ real multiplications and $L$ real additions, to detect the transmitting antenna correctly, it is necessary to traverse and search through all the antennas on the transmission end. Therefore, the computational complexity of (\ref{intdec}) becomes $(L+4)N_t$ multiplications and $(L+1)N_t$ additions.
\section{Performance Analysis}
In this section, we derive the performance of the RIS-SSK scheme under imperfect CSI, where the RIS is used to connect the Rayleigh fading channel on the BS-RIS side and the Rician fading channel on the RIS-UE side. The CPEP and UPEP expressions with the optimal ML detector are derived. Further, the corresponding ABEP expression of the RIS-SSK scheme with the imperfect CSI is obtained.
\subsection{CPEP Expression}
It is assumed that the activated transmit antenna index is $n_t$ and the detected antenna index is $\hat n_t$. By exploiting the decision rules provided in (\ref{intdec}), the CPEP can be given as
\begin{equation}\label{xdfsg0}
\begin{aligned}
P_b =& \Pr\{n_t \to \hat{n}_t|\alpha_{l,{n_t}},\hat\beta_l\} \\
= &\Pr \{|y - \sqrt{P_s}\zeta\sum_{l=1}^L \alpha_{l,{n_t}}\hat\beta_l|^2 \\&>|y-\sqrt{P_s}\zeta\sum_{l=1}^L \alpha_{l,{\hat{n}_t}}\hat\beta_l e^{-j(\theta_{l,n_t}-\theta_{l,{\hat{n}_t}})}|^2\}\\
=&\Pr\{-2\Re\{y\sqrt{P_s}\zeta\sum_{l=1}^L \alpha_{l,{n_t}}\hat\beta_l\}+|\sqrt{P_s}\zeta\sum_{l=1}^L \alpha_{l,{n_t}}\hat\beta_l|^2 \\&>-2\Re\{y\sqrt{P_s}\zeta\sum_{l=1}^L \alpha_{l,{\hat{n}_t}}\hat\beta_l e^{-j(\theta_{l,n_t}-\theta_{l,{\hat{n}_t}})}\}\\&+|\sqrt{P_s}\zeta\sum_{l=1}^L \alpha_{l,{\hat{n}_t}}\hat\beta_l e^{-j(\theta_{l,n_t}-\theta_{l,{\hat{n}_t}})}|^2\}.
\end{aligned}
\end{equation}
To simplify the representation of (\ref{xdfsg0}), let us define
\begin{equation}\label{kap10}
\eta = \sum_{l=1}^L \alpha_{l,{n_t}}\hat\beta_l, \ \ \hat\eta=\sum_{l=1}^L \alpha_{l,{\hat{n}_t}}\hat\beta_l e^{-j(\theta_{l,n_t}-\theta_{l,{\hat{n}_t}})}.
\end{equation}
Substituting (\ref{kap10}) into (\ref{xdfsg0}), the CPEP can be updated to
\begin{equation}\label{sdfggsdg0}
\begin{aligned}
P_b
=&\Pr(-2\Re\{y\sqrt{P_s}\zeta\eta\}+|\sqrt{P_s}\zeta\eta|^2\\&>-2\Re\{y\sqrt{P_s}\zeta\hat \eta\}+|\sqrt{P_s}\zeta\hat \eta|^2)\\
=&\Pr\left(2\Re\{y\sqrt{P_s}\zeta(\hat\eta-\eta)\}\right.\\&+\left.|\sqrt{P_s}\zeta\eta|^2 -|\sqrt{P_s}\zeta \hat \eta|^2>0\right).
\end{aligned}
\end{equation}
Recall that (\ref{y02}), let us define $u = \sum_{l=1}^Lg_{l,n_t}e^{j\phi_{l,n_t}}\Delta h_{l}$.\footnote{For two independent random variables $X$ and $Y$, we can obtain the expectation and variance of term $XY$ as $E(XY)=E(X)E(Y)$ and $Var(XY)=Var(X)Var(Y)+Var(X)E^2(Y)+E^2(X)Var(Y)$, respectively.}
By adopting CLT, the $u$ obeys $\mathcal{CN}(0,\sigma_e^2L)$.
In this manner, (\ref{sdfggsdg0}) can be recast as
\begin{equation}\label{sdfggsdg1}
\begin{aligned}
P_b
=&\Pr(2\Re\{(\sqrt{P_s}\zeta\eta + \sqrt{P_s(1-\zeta^2)}u + n_0)\sqrt{P_s}\zeta(\hat\eta-\eta)\}\\&+|\sqrt{P_s}\zeta\eta|^2  -|\sqrt{P_s}\zeta \hat \eta|^2>0)\\
=&\Pr(2\Re\{(\sqrt{P_s(1-\zeta^2)}u + n_0)\sqrt{P_s}\zeta(\hat\eta-\eta)\}\\&-|\sqrt{P_s}\zeta\eta|^2 -|\sqrt{P_s}\zeta \hat \eta|^2-2P_s\zeta^2\eta\hat\eta>0)\\
=&\Pr(2\Re\{(\sqrt{P_s(1-\zeta^2)}u + n_0)\sqrt{P_s}\zeta(\hat\eta-\eta)\}\\&-|\sqrt{P_s}\zeta\hat\eta-\sqrt{P_s}\zeta\eta|^2>0)\\
=&\Pr\left(D>0\right),
\end{aligned}
\end{equation}
where $D\sim\mathcal{N}(\mu_D,\sigma_D^2)$. The expectation and variance of $D$ are represented as $\mu_D=-P_s\zeta^2|\hat\eta-\eta|^2$ and $\sigma_D^2={2(N_0+P_s(1-\zeta^2)\sigma_e^2L)}$, respectively.
In this respect, the (\ref{sdfggsdg1}) can be evaluated as
\begin{equation}\label{enu01}
\begin{aligned}
P_b
=&\Pr(-\mu_D/\sigma_D)=Q\left(\sqrt{\frac{P_s\zeta^2|\hat\eta-\eta|^2}{2(N_0+P_s(1-\zeta^2)\sigma_e^2L)}}\right).
\end{aligned}
\end{equation}
\subsection{UPEP Expression}
By employing (\ref{kap10}), the term of $\eta -\hat \eta$ can be written as
\begin{equation}\label{enu02}
\eta -\hat \eta = \sum_{l=1}^L \hat\beta_l \left(\alpha_{l,{{n}_t}}- \alpha_{l,{\hat{n}_t}} e^{-j\omega}\right).
\end{equation}
where $\omega=\theta_{l,n_t}-\theta_{l,{\hat{n}_t}}$.
Since $\theta_{l,n_t}$ and $\theta_{l,{\hat{n}_t}}$ both independently and uniformly distributed in $(0,2\pi)$, then the PDF of $\omega$ can be given as follows:
\begin{equation}\label{eqphix}
f_{\omega}(x)=\left\{
\begin{aligned}
&\frac{1}{2\pi}(1+\frac{x}{2\pi}),\ \ \  x \in [-2\pi,0),\\
&\frac{1}{2\pi}(1-\frac{x}{2\pi}), \ \ \ x \in [0,2\pi).\\
\end{aligned}
\right.
\end{equation}
In this manner, the $\alpha_{l,{\hat{n}_t}} e^{-j\omega}$ in (\ref{enu02}) can be calculated as
\begin{equation}
\alpha_{l,{\hat{n}_t}} e^{-j\omega}=\alpha_{l,{\hat{n}_t}}\cos\omega-j\alpha_{l,{\hat{n}_t}}\sin\omega.
\end{equation}
Since the symmetry of cosine and sine function, we have
\begin{subequations}
\begin{align}
&E[\alpha_{l,{\hat{n}_t}} e^{-j\omega}] = 0, \\ &Var[(\alpha_{l,{\hat{n}_t}} e^{-j\omega})_\Re] = 1/2, \\ &Var[(\alpha_{l,{\hat{n}_t}} e^{-j\omega})_\Im] = 1/2.
\end{align}
\end{subequations}
Note that subsequent section requires fitting the distribution using the CLT, the expectation and variance of each variable need to be obtained.
It is known that the real and imaginary parts are two independent parts of each other, thus the variance of $\alpha_{l,{\hat{n}_t}} e^{-j\omega}$ is $Var[(\alpha_{l,{\hat{n}_t}} e^{-j\omega})] = 1$.

After some simple algebraic operations, we can derive the mean and variance of $\alpha_{l,{{n}_t}}- \alpha_{l,{\hat{n}_t}} e^{-j\omega}$ in (\ref{enu02}) as
\begin{subequations}
\begin{align}
&E(\alpha_{l,{{n}_t}}- \alpha_{l,{\hat{n}_t}} e^{-j\omega}) = \frac{\sqrt{\pi}}{2},\\
&Var(\alpha_{l,{{n}_t}}- \alpha_{l,{\hat{n}_t}} e^{-j\omega}) = \frac{8-\pi}{4}.
\end{align}
\end{subequations}
Further, the mean and variance of $\beta_l \left(\alpha_{l,{{n}_t}}- \alpha_{l,{\hat{n}_t}} e^{-j\omega}\right)$ in (\ref{enu02}) can be respectively expressed as
\begin{subequations}
\begin{align}
&E[\hat\beta_l (\alpha_{l,{{n}_t}}- \alpha_{l,{\hat{n}_t}} e^{-j\omega})] = \frac{\sqrt{\pi}}{2}E(\hat\beta_l),\\
&Var[\hat\beta_l (\alpha_{l,{{n}_t}}- \alpha_{l,{\hat{n}_t}} e^{-j\omega})] =2-\frac{\pi}{4}E^2(\hat\beta_l).
\end{align}
\end{subequations}
Due to the fact that each reflector element of a RIS is independent of each other, it is difficult to directly obtain an accurate PDF for the sum of the reflective elements. To address this issue, we can use the CLT to approximate the PDF as a real Gaussian distribution. Consequently, the corresponding mean and variance can be respectively expressed as
\begin{equation}\label{musigma}
\mu = \frac{\sqrt{\pi}LE(\hat\beta_l)}{2},\ \
\sigma^2 = \frac{L[8-\pi E^2(\hat\beta_l)]}{4}.
\end{equation}
Based on (\ref{enu01}) and (\ref{musigma}), the UPEP of the proposed scheme can be calculated as
\begin{equation}\label{q1}
\begin{aligned}
    \bar P_b &= \int_0^\infty Q\left(\sqrt{\frac{\rho\zeta^2x}{2(1+\rho(1-\zeta^2)\sigma_e^2L)}}\right)f(x)dx,
\end{aligned}
\end{equation}
where $x = |\eta-\hat\eta|^2$, $\rho={P_s}/{N_0}$ stands for SNR, and $f(x)$ denotes the PDF of $x$ variable.
According to  \cite{zhu2023qua}, we have
\begin{equation}\label{qapr}
Q(x) \approx \frac{1}{12}\exp\left(-\frac{x^2}{2}\right)+\frac{1}{4}\exp\left(-\frac{2x^2}{3}\right).
\end{equation}
Substituting (\ref{qapr}) into (\ref{q1}), the UPEP can be reformulated as
\begin{equation}\label{qapr01}
\begin{aligned}
\bar P_b
\approx &\frac{1}{12}\int_0^\infty \exp\left(-{\frac{\rho\zeta^2x}{4(1+\rho(1-\zeta^2)\sigma_e^2L)}}\right)f(x)dx \\
& +\frac{1}{4}\int_0^\infty\exp\left(-{\frac{\rho\zeta^2x}{3(1+\rho(1-\zeta^2)\sigma_e^2L)}}\right)f(x)dx.
\end{aligned}
\end{equation}
Since the variable $x$ represents the non-central chi-square distribution with one degree of freedom, its PDF form is very cumbersome. To facilitate the address (\ref{qapr01}), we resort to \cite{li2021space} to obtain its moment-generating function  as
\begin{equation}\label{qapr11}
M_X(s)=\frac{1}{\sqrt{1-2s\sigma^2}}\exp\left(\frac{\mu^2s}{1-2s\sigma^2}\right).
\end{equation}
Substituting (\ref{qapr11}) into (\ref{qapr01}), we obtain the closed-form expression of UPEP as
\begin{equation}\label{qapr1}
\begin{aligned}
\bar P_b
=&\frac{1}{12}\sqrt{\frac{{2(1+\rho(1-\zeta^2)\sigma_e^2L)}}{{2(1+\rho(1-\zeta^2)\sigma_e^2L)}+{{\rho\zeta^2\sigma^2}}}}\\&\times\exp\left({\frac{-\mu^2\rho\zeta^2}{{4(1+\rho(1-\zeta^2)\sigma_e^2L)}+{{2\rho\zeta^2\sigma^2}}}}\right) \\
 +&\frac{1}{4}\sqrt{\frac{3(1+\rho(1-\zeta^2)\sigma_e^2L)}{3(1+\rho(1-\zeta^2)\sigma_e^2L)+2\sigma^2\rho\zeta^2}}\\&\times\exp\left(\frac{-\mu^2\rho\zeta^2}{3(1+\rho(1-\zeta^2)\sigma_e^2L)+2\sigma^2\rho\zeta^2}\right).
\end{aligned}
\end{equation}
After some manipulations, (\ref{qapr1}) can rewritten as
\begin{equation}\label{qapr2}
\begin{aligned}
\bar P_b
=&\frac{1}{12}\sqrt{\frac{{2(1+\rho\sigma_e^4L)}}{{2(1+\rho\sigma_e^4L)}+{{\rho\sigma^2}}}}\exp\left({\frac{-\mu^2\rho}{{4(1+\rho\sigma_e^4L)}+{{2\rho\sigma^2}}}}\right) \\
 +&\frac{1}{4}\sqrt{\frac{3(1+\rho\sigma_e^4L)}{3(1+\rho\sigma_e^4L)+2\sigma^2\rho}}\exp\left(\frac{-\mu^2\rho}{3(1+\rho\sigma_e^4L)+2\sigma^2\rho}\right).
\end{aligned}
\end{equation}
\subsection{Asymptotic UPEP}
To provide a better demonstration of the impact of the channel estimation error parameters on the performance of the considered RIS-SSK system, we evaluate the performance of this system with respect to the high SNR region. Here, the asymptotic expression for the UPEP can be calculated as
\begin{equation}\label{eqasy}
\begin{aligned}
&\bar P_a= \lim\limits_{\rho \to \infty}\bar P_b=\\
&\frac{1}{6}\sqrt{\frac{2\sigma_e^4}{8\sigma_e^4+8-\pi E^2(\hat\beta_l)}}\exp\left({\frac{-\pi LE^2(\hat\beta_l)}{{16\sigma_e^4}+{16-2\pi E^2(\hat\beta_l)}}}\right) \\
& +\frac{1}{4}\sqrt{\frac{6\sigma_e^4}{6\sigma_e^4+8-\pi E^2(\hat\beta_l)}}\exp\left(\frac{-\pi L E^2(\hat\beta_l)}{6\sigma_e^4+8-\pi E^2(\hat\beta_l)}\right).
\end{aligned}
\end{equation}

\subsection{ABEP}
It is worth noting that ABEP is equal to UPEP when $N_t$ is two, while ABEP is the union upper bound of the scheme if $N_t$ is greater than two. Consequently, the ABEP of the RIS-SSK scheme can be characterized as \cite{canbilen2022on}
\begin{equation}
ABEP \leq \frac{1}{\log_2N_t}\sum_{\hat n_t=1}^{N_t}\sum_{n_t\neq \hat n_t}^{N_t} \bar P_i N(\hat n_t\to n_t), \ \ \ i \in\{a,b\}
\end{equation}
where $N(\hat n_t\to n_t)$ indicates the number of error bits between the true transmit antenna index $n_t$ and the decoded judgment obtained antenna index $\hat n_t$.
\section{Simulation and Analytical Results}
In this section, we explore the error performance of the proposed scheme under imperfect CSI via Monte Carlo simulation. The simulation involves generating a random data sequence and transmitting it to the receiver via RIS reflection after modulation.  In simulations, the ABEP value corresponding to each SNR is generated $1\times10^6$ times and then the average value is calculated, which is used to validate the analytical derivations. Unless otherwise specified, $N_t$ and $N_r$ are set to 2 and 1, respectively, and RIS is a square array. Additionally, the impact of any large-scale path loss is ignored as it is already implicit in the received SNR.

\begin{figure}[t]
\centering
\includegraphics[width=6.5cm]{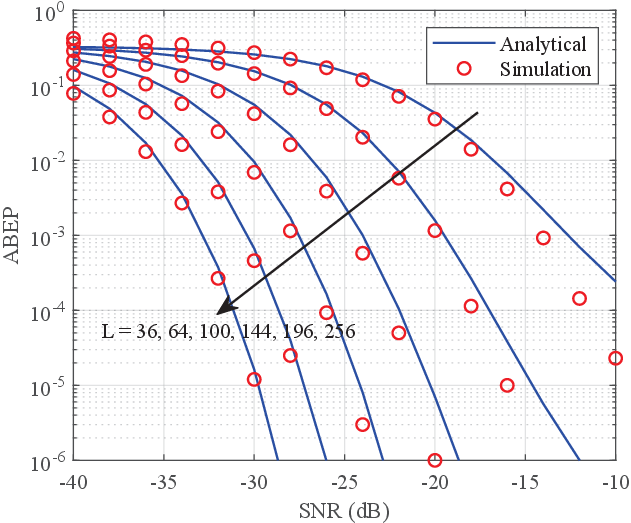}
\caption{\small{Validation of analytical ABEP via the simulation.}}
\vspace{-10pt}
\label{verfsa}
\end{figure}

In Fig. \ref{verfsa}, we plot the ABEP performance of the RIS-SSK scheme under imperfect CSI, where the Rician factor $\kappa$ the error estimation parameters $\sigma_e^2$ are set as 3 dB and 0.1, respectively.
It is worth noting that the simulation results in (\ref{intdec}) and the closed-form expression in (\ref{qapr2}) start to agree with the variation of SNR when $L$ is not less than 144 since the CLT requires at least two orders of magnitude.
In addition, we observe that the simulation value coincides almost perfectly with the analytical result in the case of $L=256$, which further validates the correctness of the derived result and shows that the gap between (\ref{qapr01}) and the real value is very small and almost negligible.
\begin{figure}[t]
\centering
\includegraphics[width=6.5cm]{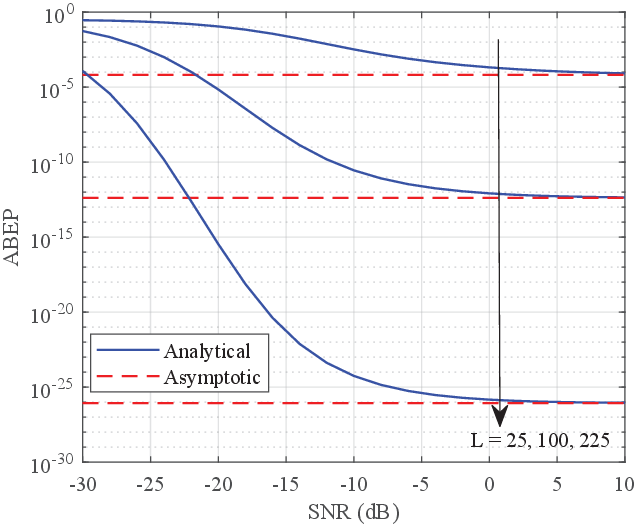}
\caption{\small{Validation of asymptotic ABEP of the RIS-SSK scheme.}}
\vspace{-10pt}
\label{intasy}
\end{figure}

\begin{figure}[t]
\centering
\includegraphics[width=6.5cm]{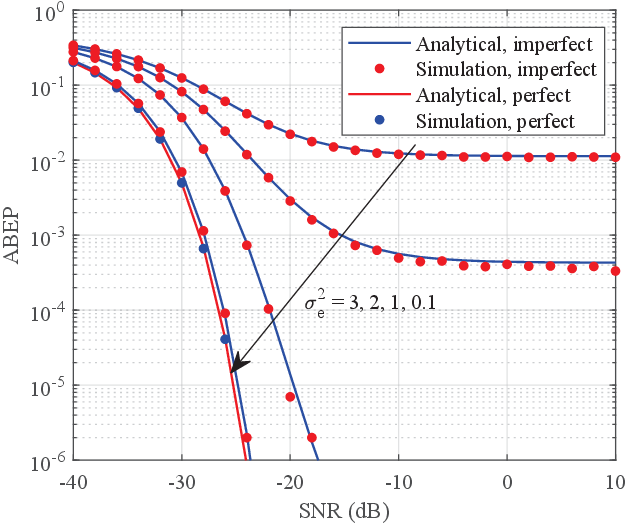}
\caption{\small{Impact of estimation error $\sigma_e^2$ on ABEP.}}
\vspace{-10pt}
\label{asy}
\end{figure}

\begin{figure}[t]
\centering
\includegraphics[width=6.5cm]{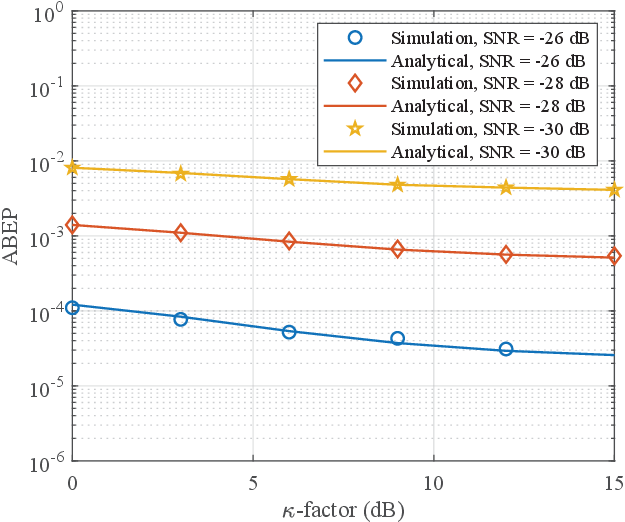}
\caption{\small{Impact of Rician factor $\kappa$ on ABEP.}}
\vspace{-10pt}
\label{kapver}
\end{figure}

In Fig. \ref{intasy}, we validate the correctness of the developed asymptotic ABEP, where the Rician factor $\kappa$ and the error estimation parameter $\sigma_e^2$ are setup at 3 dB and 0.1, respectively.
Specifically, in Fig. \ref{intasy}, the analytical and asymptotic ABEP curves are plotted, which are generated based on (\ref{qapr2}) and (\ref{eqasy}), respectively.
It can be observed that as the SNR increases, there is not only a significant performance degradation but also an error floor, which is caused by the channel estimation error. As the number of reflective elements $L$ increases, the ABEP of the scheme decreases accordingly. However, for high SNR regions with imperfect CSI, this is not the case, since the dominant noise is no longer AWGN, but originates from channel estimation errors.

In Fig. \ref{asy}, the Monte Carlo simulation results and analytical curves of the RIS-SSK scheme with $\kappa$ = 3 dB and $L=144$ are given.
Note that the error of channel estimation is set as fixed values $\sigma_e^2 = 3,2,1,0.1$, respectively, that is, the correlation coefficients are $\zeta = 0.500,0.5774,0.7071,0.9535$.
As a reference, the corresponding ABEP with the perfect CSI is also shown with a dashed line for the RIS-SSK scheme.
From Fig. \ref{asy}, it can be seen that the RIS-SSK scheme has good anti-noise performance. According to \cite{basar2012per}, $\sigma_e^2$ is taken to be much less than 0.1 to approach the value of perfect CSI, while in this figure, it is taken to 0.1 to be very close to the ABEP with perfect CSI.

In Fig. \ref{kapver}, we exhibit the performance impact of the LoS path of the reflection channel on the RIS-SSK scheme with imperfect CSI, where the number of reflecting elements is 144 and the estimation error variance of each reflecting element up to the UE is 0.1.
From Fig. \ref{kapver}, simulation and analytical values match very well. The error can be reduced by increasing the number of simulations.
Moreover, the simulation and analytical values exhibit a strong correlation. To decrease the error, one can increase the number of simulations or increasing the SNR values.
A higher Rician factor indicates a stronger reflected LoS path signal energy, which in turn results in a higher quality of the received signal on the UE side. As a result, the ABEP performance can be improved.
Additionally, it is also found that the ABEP performance of RIS-SSK in the imperfect CSI case is enhanced as the SNR increases.

\section{Conclusion}
In this paper, the performance of RIS-SSK with imperfect channel estimation is analyzed, where BS-RIS channel suffers from Rayleigh fading and RIS-UE channel follows the Rician fading. Based on the ML detector, we derived the CPEP expression and the PDF of the composite channel with a non-central chi-square distribution with one degree of freedom.
After that, we derive the  closed-form expression and asymptotic expression of ABEP under the RIS-SSK scheme with impact CSI.
Finally, all the analytical derivations are verified by Monte Carlo simulation and it is found that the ABEP values obtained are closer to the true results as the channel estimation error is smaller or the Rician factor is larger.

\end{document}